\documentstyle[12pt,epsf]{article}
\newcommand{\lsim}{
\makebox[3.0ex][c]{\raisebox{ 0.4ex}{\large $<$}}\hspace*{-3.0ex}
\makebox[3.0ex][c]{\raisebox{-0.5ex}{\large ${\sim}$}}}

\begin{document}

\begin{center}

{\large\bf Localized plasmons in point contacts\\[5mm]
}

Henrik Bruus$^a$ and Karsten Flensberg$^b$ \\
{\em \mbox{}$^a$Niels Bohr Institute, Blegdamsvej 17, DK-2100 Copenhagen}\\
{\em \mbox{}$^b$Danish Institute of Fundamental Metrology, 
 Anker Engelunds Vej 1, DK-2800 Lyngby}\\[2mm]
(Semicond.\ Sci.\ Technol.\ {\bf 13}, A30, 1998).
\end{center}

\vspace*{5mm}
\noindent {\bf Abstract}\\
Using a hydrodynamic model of the electron fluid in a point
contact geometry we show that localized plasmons are likely to exist
near the constriction. We attempt to relate these plasmons with the
recent experimental observation of deviations of the quantum point
contact conductance from ideal integer quantization. As a function of 
temperature this deviation exhibits an activated behavior, 
$\exp(-T_a/T)$, with a density dependent activation temperature $T_a$
of the order of 2 K. We suggest that $T_a$ can be identified with the
energy needed to excite localized plasmons, and we discuss the
conductance deviations in terms of a simple theoretical model
involving quasiparticle lifetime broadening due to coupling to the
localized plasmons. 

\vspace*{5mm}
\noindent {\bf Introduction}\\
The quantized conduction through a narrow point contact is one of the
key effects in mesoscopic physics, the quantum point contact remains
an important testing ground for the description of mesoscopic
phenomena. Recently, significant deviations from the
Landauer-B\"{u}ttiker theory have been observed in quantum
point contacts in the temperature dependence of the conductance
quantization \cite{Tarucha95,Yacoby96} and as a so called ``0.7''
structure or quasi plateau,  appearing around $0.7$ times the
conductance quantum $2 e^{2}/h$ \cite{Thomas95}. Invoking a Luttinger
liquid approach \cite{Kane92} the deviations have been discussed in
terms of interaction effects \cite{Kawabata96,Shimizu96,Oreg96}.
However, firm conclusions have been difficult to
obtain partly due to the narrow temperature range (0.1~K - 4~K) in
which the effect can be studied in conventional split gate quantum
point contacts, where relatively close lying one-dimensional subbands
are formed. 

An important progress was provided with the appearance of strongly
confined GaAs quantum point contacts using a combination of shallow
etching and a top gate \cite{Kristensen98a}. In these new samples the
conduction quantization can be followed up to around 30~K. In a
subsequent work \cite{Kristensen98b} these samples were used to study
the temperature dependence deviations from perfect conductance
quantization. At low temperature ($\sim$ 0.05~K) almost ideal
quantized conductance is observed for the first conduction plateau,
but deviations develop as the temperature is increased. The enlarged
temperature range allowed for the observation of activated temperature
dependence of these deviations: $\delta G(T) \propto
\exp(-T_a/T)$. Furthermore, by changing the top gate it was
found that $T_a$ increases with increasing density. An explanation
could not be found using the standard single particle picture, and 
in the brief theory section of Ref.~\cite{Kristensen98b} we
therefore suggested to include collective effects through plasmons. 
In short, we identified $T_a$ as the energy
needed to excite localized plasmons, and we discussed the conductance
in terms of a simple theoretical model involving the additional effect
of electrons scattering off the localized plasmons. In the present
theoretical work we elaborate on that idea. In a point contact the
charge is of course depleted. In order to study the collective
excitations of such a system, we can approach it from two limits: 1)
starting from a homogeneous electron liquid which is 0, or 2)
starting with two spatially separated liquids. Below we follow the
first approach, and we argue from a hydrodynamic model that
localized plasmons may exist in realistic situations.

\vspace*{5mm} 
\noindent {\bf Plasmons of a homogeneous electron liquid in a cylinder}\\
Following Fetter \cite{Fetter85} we use a hydrodynamic model of a
weakly damped, compressible charged electron fluid placed in a rigid,
neutralizing positive background set to $+en_0$. The electron density is
written as $n_0+n$, where $n$ is a small perturbation, and the
electronic velocity field is denoted ${\bf v}$. Finally, we include
the electrostatic potential $\Phi$ and neglect radiation effects. The
basic equations for the system are the linearized versions of the
continuity equation and of Euler's and Poisson's
equations\cite{Fetter85}: 
\begin{eqnarray} 
\label{continuity}
\partial_t n & = & - n_0 {\bf \nabla} \!\cdot\! {\bf v}\\
\label{Euler}
\partial_t {\bf v} & = & - \frac{s^2}{n_0} {\bf \nabla}n
 + \frac{e}{m} {\bf \nabla}\Phi\\
\label{Poisson}
{\bf \nabla}^2\Phi & = & \frac{en}{\epsilon} \; . 
\end{eqnarray}
Here $s = (\partial P/\partial n)/m = \sqrt{3/5} \: v_F$ is the sound
velocity of the liquid. Combining Eqs.~(\ref{Euler})-(\ref{Poisson})
and introducing the plasma frequency $\omega_p = e^2 n_0/m\epsilon$ we
obtain a wave equation for $n$:
\begin{equation} \label{n_waveeq}
-s^2 {\bf \nabla}^2 n + \omega_p^2 n = \omega^2 n.
\end{equation}
For the case of a homogeneous electron liquid confined in a cylinder of
radius $R$ we let all fields have the dependence $f(r,\theta,z,t) =
f_l(r) \exp[i(l\theta + qz - \omega t)]$, with $f_l$ being a
Bessel function. Outside the cylinder $\Phi(r)$ must decay and
fulfill Eq.~(\ref{Poisson}) with $n=0$, and so $\Phi_>(r) \propto K_l(q
r)$. Inside the cylinder, $\Phi_<(r)$ can be either decaying, as
$I_l(\kappa r)$, or oscillating, as $J_l(k r)$. The lowest lying
modes are the decaying ones reminiscent of surface plasmons. The
oscillation frequency $\omega$ is found by enforcing the boundary
conditions that $\Phi(r)$ and its derivative are continuous at $r=R$,
and that the normal component $v_r$ of the velocity vanishes at
the surface. The solution is 
\begin{equation} \label{omega_homogen_cyl}
\omega^2 = qRI'_l(qR) \left( K_l(qR) - 
\frac{q}{\kappa} \frac{K'_l(qR)}{I'_l(\kappa R)} I'_l(\kappa R)
\right) \omega^2_p  \;\;
_{\stackrel{\displaystyle 
\longrightarrow}{\scriptstyle q \rightarrow 0}} \; \;
\frac{e^2 n^{1D}}{2 \pi m \epsilon} \;
q^2 \; \ln(\frac{2}{qR}),
\end{equation}
where we also have given the 1D-limit arising as $q \rightarrow 0$.

\vspace*{5mm} 
\noindent {\bf Plasmons of an inhomogeneous electron liquid in a squeezed
elliptical cylinder}\\
Next, to approach the point contact geometry we introduce two
perturbations. First, the cylinder containing the inhomogeneous
electron liquid is squeezed geometrically in a region of length $2
\Lambda$ around $z=0$, {\it i.e.\/}\ the radius becomes a function of
$z$, say for example $R(z)= R_0-\delta R [1+\cos(\pi z/\Lambda)]
\Theta(\Lambda-|z|)$. Similarly, a static $z$ dependent dip is imposed
on the positive background charge density $n_0$ inside the squeezed
cylinder, say $n_0(z) =  n_0 - \delta n [1+\cos(\pi z/L)]
\Theta(L-|z|)$.

In the adiabatic limit where derivatives of $R(z)$ and $n_0(z)$ are
neglected, the wave equation Eq.~(\ref{n_waveeq}) remains separable in
cylindrical coordinates, and we make the {\it ansatz} $n(r,\theta,z) =
J_l(\kappa r) g(z) \exp[i(l\theta - \omega t)]$, where $J_l$ is a
Bessel function and $g(z)$ an arbitrary function to be determined.
The boundary condition $v_r(R(z)) = 0$ translates into a Neumann
boundary condition $J'_l(\kappa R(z)) = 0$ and consequently
the ``wavenumber'' $\kappa$ becomes a function of $z$,
$\kappa = \kappa_{nl}(z) = \tilde{\gamma}_{nl}/R(z)$, with
$\tilde{\gamma}_{nl}$ being the $n$'th root of $J'_l(x)$. Furthermore,
$\omega_p^2$ also becomes a function of $z$, since $\omega_p^2(z) =
e^2 n_0(z)/m \epsilon$, and similarly for the sound velocity, $s =
s(z) \propto n_0(z)^{1/3}$. As a consequence the wave equation
Eq~(\ref{n_waveeq}) for $n$ is changed into the following
eigenfunction equation for $g(z)$: 

\begin{equation} \label{g_waveeq}
-s(z)^2\partial_z^2 g(z) +
[s(z)^2 \kappa^2_{nl}(z) + \omega^2_p(z)] \: g(z) = \omega^2 g(z). 
\end{equation}

This is equivalent to Schr\"{o}dinger's equation (with a position
dependent mass) as seen by the identifications $s^2 \leftrightarrow
\hbar^2/2m$ and $s(z)^2 \kappa^2_{nl}(z) + \omega^2_p(z)
\leftrightarrow V(z)$. Since $\omega^2_p(0) < \omega^2_p(\pm\infty)$
bound states, {\it i.e.\/}\ localized plasmons, may exist.  
The ``effective potential'' $V(z)$ is a sum of two terms; one,
$\omega^2_p$ is dipping down on the length scale L, the other,
$s^2 \kappa_{nl}^2$, is peaking on the length scale
$\Lambda$. Depending on the relative strengths, shapes and
length scales of the two terms the effective potential will appear
rather differently. However, for realistic parameters, where the
density variation dominates, we conclude that localized plasmons may
exist in the squeezed, inhomogeneous cylindrical electron
liquid as shown in Fig.~1.

The previous considerations dealt with a cylindrical geometry, but it
is not difficult to approach the 2D case. The trick is simply to use
elliptical coordinates $(u,v,z)$ defined by $(x,y,z) = (\eta \cosh(u)
\cos(v), \eta \sinh(u) \sin(v), z)$. The parameter $\eta$ relates to
the eccentricity of the ellipse. With these coordinates the wave
equation separates as before. Instead of trigonometric functions of
the angle $\theta$ we now obtain the Mathieu functions of the
generalized angular variable $v$, and instead of Bessel functions we
obtain the modified Mathieu functions of the generalized radial
coordinate $u$. By letting the eccentricity $\eta$ tend to infinity we
end up with a 2D geometry close to the one realized in the quantum
point contact experiments. The conclusions obtained for the circular
cylinder can be restated for the elliptic cylinder, and thus localized
plasmons are expected to exists in or near the constriction region of
quantum point contacts.

\begin{figure}[h]
\epsfysize=50mm \centerline{\epsfbox{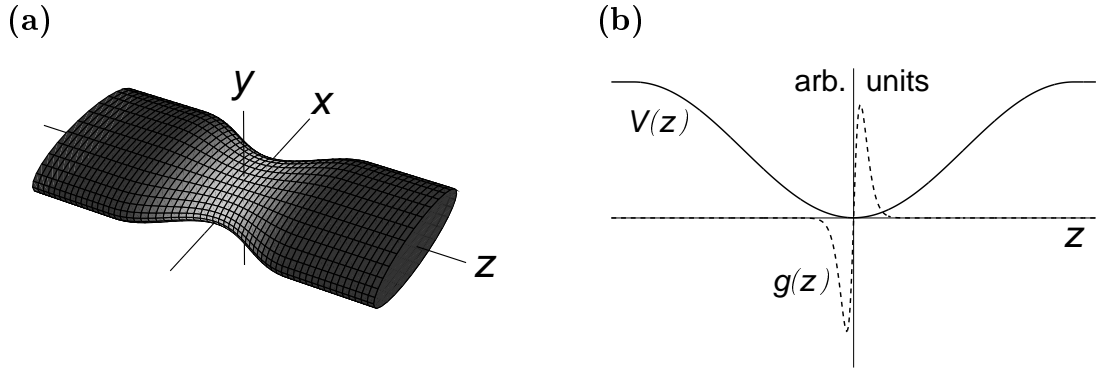}}
\caption{
(a) The squeezed elliptical cylinder. High and low densities are
represented by dark and light shadings respectively. (b) The
effective potential $V(z)$ (full line) is determined by the
parameter values of Ref.~\protect\cite{Kristensen98b}. For this
potential a solution of the wave equation Eq.~(\protect\ref{g_waveeq})
for $g(z)$ is found numerically (dashed line). The solution
represents a localized plasmon with an energy of the order 10~K.
}
\end{figure}

\vspace*{5mm} 
\noindent {\bf Plasmon damping}\\
So far we have treated only the undamped case. In real systems the
collective plasmons are damped through their interaction with
individual electron-hole pairs, the so called Landau damping. This
effect could be simulated by adding a damping term $-{\bf v}/\tau$ to
the right hand side of the Euler equation Eq.~(\ref{Euler}). Instead
we will leave the classical level of description and continue with a
microscopic quantum treatment. The classical level is adequate for
demonstrating the existence of the collective (almost classical)
plasmon excitations, but fails when it comes to single particle
effects. 

The point contact can be approximated by a 1D region, the
constriction, connected at each end to 2D regions, the contacts.
For this 2D-1D-2D model of the point contact we can estimate the
frequency of the confined plasmon using our insight from the classical
calculations: we calculate the dispersion relation for an infinite
1D-wire and insert the wave vector $q_c = 2\pi /L$, L being the
length of the constriction and hence related to the size of the localized
plasmon. The long wave length limit of the dispersion relation found
by a RPA calculation is
\begin{equation} \label{omega_1D}
\omega^{1D}_L =  \sqrt{v_f^2 +  
\frac{\gamma e^2 n^{1D} }{4 \pi \varepsilon m^*}} \; q_c.
\end{equation}
Note how the second term under the square root resembles the classical
result of Eq.~(\ref{omega_homogen_cyl}). In Ref.~\cite{Kristensen98b}
we used this formula successfully to fit the measured activation
temperatures mentioned in the introduction. 

The confined 1D-plasmons will be Landau-damped through their coupling
to the 2D-contacts outside the constriction. Inserting $q_c$ in the
RPA expression for the polarizability $\chi^{2D}$ we obtain 
the following rough estimate of the lifetime $\tau_p^{-1}/\omega^{1D}_L$
of the 1D-plasmon coupled to the 2D-contacts of the 2D-1D-2D model:

\begin{equation} \label{damping}
\frac{\tau_p^{-1}}{\omega^{1D}_L} \approx V^{2D} {\rm Im}\chi^{2D} 
\approx \frac{2\pi e_0^2}{q_c} 
\frac{m^* \omega^{1D}_L}{2\pi \hbar^2 v_F^{2D}q_c} \lsim 1.
\end{equation}
The plasmons are seen to be damped, but not over-damped.

\vspace*{5mm} 
\noindent {\bf Quasiparticle lifetime}\\
In the Landauer-B\"uttiker formalism the conductance is given by
single particle properties. Once a particle is launched in a given
channel of the injecting lead the transmission probability amplitudes
are governed by the elastic scattering matrix of the system. For a
quasiparticle with a finite lifetime it is possible that a particle
will decay before completing its traversal of the system. We propose
that the observed deviation from perfect quantized conductance is
indeed due to the finite lifetime of the quasiparticles. Furthermore
we suggest that the main contribution restricting the lifetime comes
from scattering against the localized plasmons. As demonstrated above,
the localized plasmons provide a well defined finite energy $\hbar
\omega_L$. Through the Coulomb interaction the electrons will
scatter against the plasmons and hence the quasiparticle lifetime and
the transmission properties are affected. The resulting lifetime and
additional resistance is expected to exhibit an activated behavior,
$\tau^{-1} \propto \exp(-T_a/T)$, since a finite energy is needed to
excite the localized plasmon. We are thus lead to identify the
activation temperature with the energy of the localized plasmon: $T_a
= \hbar \omega_L/k_B$.

\vspace*{5mm} 
\noindent {\bf Conclusion}\\
Using a hydrodynamic description of the electron fluid, we have shown
that localized plasmons with a frequency $\omega_L$ are likely to exist
near the constriction of a point contact. We have sketched how
a more complete microscopic quantum calculation may account for a  
quasiparticle lifetime broadening $\tau^{-1}$ with a thermal
activation behavior $\tau^{-1} \propto \exp(-\hbar \omega_L/k_BT)$. 
We relate this broadening with conductance and are lead to identify
the recently measured activation temperature $T_a$ for conductance
deviations with the frequency $\omega_L$ of the localized plasmon.

\vspace*{5mm} 
\noindent {\bf Acknowledgements}\\
It is a pleasure to thank our experimental colleagues Anders
Kristensen, Poul Erik Lindelof, and Jesper Nyg\aa rd, at the \O rsted 
Laboratory of the Niels Bohr Institute, for many stimulating
discussions. HB is supported by the Danish Natural Science
Research Council through Ole R\o mer Grant no.\ 9600548. 

\pagebreak


\begin{thebibliography}{99}

\bibitem{Tarucha95} Tarucha S, Honda T and Saku T 1995 {\em
Solid State Comm.} {\bf 94}, 413

\bibitem{Yacoby96} Yacoby A, Stormer H L, Wingreen N S, Pfeiffer L N,
Baldwin K W, and West K W 1996 {\em Phys.\ Rev.\ Lett.} {\bf 77}, 
4612

\bibitem{Thomas95} Thomas K J, Simmons M Y, Nicholls J T,
Mace D R, Pepper M and Ritchie D A 1995 {\em Appl. Phys. Lett.} {\bf
67}, 109 

\bibitem{Kane92} Kane C L and Fisher M P A 1992 {\em Phys.\ Rev.\ B}
{\bf 46} 15233

\bibitem{Kawabata96} Kawabarta A 1996 {\em J.\ Phys.\ Soc.\ Jap.} {\bf
65} 30

\bibitem{Shimizu96} Shimizu A 1996  {\em J.\ Phys.\ Soc.\ Jap.} {\bf
65} 1162 

\bibitem{Oreg96} Oreg Y and Finkelstein A M 1996 {\em Phys.\ Rev.\ B}
{\bf 54} 14265

\bibitem{Kristensen98a} Kristensen A 1998 {\em et al.} 
{\em J.\ Appl.\ Phys.} {\bf 83} (in press)

\bibitem{Kristensen98b} Kristensen A 1998 {\em et al.} 
{\em Physica B} (in press), cond-mat/9807277

\bibitem{Fetter85}  Fetter A L {\em Phys.\ Rev.\ B} {\bf 32} 7676


\end{thebibliography}
\end{document}